\documentclass[english]{article}
\usepackage[T1]{fontenc}
\usepackage{subfigure}
\usepackage{graphicx}
\usepackage{amssymb}

\makeatletter

 \newcommand{\lyxaddress}[1]{
   \par {\raggedright #1 
   \vspace{1.4em}
   \noindent\par}
 }

\usepackage{babel}
\makeatother
\begin{document}

\title{Geodesics around Weyl-Bach's Ring Solution}

\author{L.A. D'Afonseca%
\footnote{akiles@thorus-scisoft.com.br%
}, P.S. Letelier%
\footnote{letelier@ime.unicamp.br%
} and S.R. Oliveira%
\footnote{samuel@ime.unicamp.br%
}}

\maketitle

\lyxaddress{Departamento de Matem\'{a}tica Aplicada, IMECC\\
Universidade Estadual de Campinas, UNICAMP\\
13083-970 Campinas. S.P., Brazil }

\begin{abstract}
{\normalsize We explore some of the gravitational features of a uniform
infinitesimal ring both in the Newtonian potential theory and in General
Relativity. We use a spacetime associated to a Weyl static solution
of the vacuum Einstein's equations with ring like singularity. The
Newtonian motion for a test particle in the gravitational field of
the ring is studied and compared with the corresponding geodesic motion
in the given spacetime. We have found a relativistic peculiar attraction:
free falling particle geodesics are lead to the inner rim but never
hit the ring.}{\normalsize \par}

\noindent {\normalsize PACS numbers: 02.40.-k, 04.20.-q, 04.20.Jb,
04.40.-b, 04.25.-g}{\normalsize \par}
\end{abstract}

\section{Introduction}

The astrophysical and elementary particle physics importance of ring
like configurations is evident: there are several ring like structures
as for example in galaxies \cite{GalacticRings} and planets; and
the lowest energy state of a closed string is a ring. However, there
are only few solutions of the Einstein's field equation that represent
the gravitational field of a ring. With somehow constrained hypothesis
of static configuration we mention the Weyl-Bach solution which is
the general relativistic analog of a Newtonian ring of constant density
\cite{Weyl(1922)}, given in terms of elliptic functions. The ring
is not a simple line source \cite{Israel66} as it will be clear from
the strange effect they have on the particle motion. The solution
is asymptotically flat but the outer communication region is not simply
connected.

We should mention several studies of self-gravitating Newtonian rotating
rings. The rotation is of primordial importance to the ring's stability
(and instabilities as well) but it has no effect on the Newtonian
potential and even in General Relativity its (magnetic part of the
curvature) contribution to geodesics is usually weaker than the sole
static (electric part of the curvature) one.

The purpose of this paper is to study some properties of the static
ring solution due to Weyl-Bach \cite{Weyl(1922)} using geodesics.
In particular we shall be interested in the attractive or apparent
repulsive character of the ring singularity as well as its directional
feature. Static axially symmetric solutions of the Einstein's equations
usually are characterized by the presence of string like singularities
(conic singularities) and its higher dimensions generalizations \cite{Letelier(1998)}.
In some cases, these singularities arise as supporting devices of
an otherwise dynamical configuration of masses. 

The ring has a different attraction at its rim. Free falling particles
are led to the inside part of the Weyl-Bach ring. This might be linked
to the different tension needed to keep the ring static. Besides that
the Weyl coordinates usually compact singularities and event horizon
into a lower dimensional region so that the physical distances are
not so evident. Of course the ring is black since the lapse function
vanishes at the ring and the spacetime is static. But the curvature
invariants diverge there as well.

In Sect. \ref{DSE} we present the equations to be solved. In subsect.
\ref{GWS} we present a summary of the main expressions associated
with the static axially symmetric spacetimes solutions of the Einstein's
field equations and the geodesic equations for test particles evolving
in these spacetimes. Then, subsect. \ref{NMASP} shows the associated
Newtonian equation of motion for comparison. At last, expected behavior
for axial and plane motions is given in subsect. \ref{sub:Axial-and-Equatorial}.

Next in Sect. \ref{RP} we show the potential associated to the ring.
We paid special attention to the motion of test particles: The Newtonian
motion (section \ref{sec:Newtonian-Motion}) and geodesic motion (section
\ref{sec:Geodesics-in-Weyl-Bach}) in the spacetimes associated to
the Weyl solution are considered. Most of the motion is not trivial.

We think it is important to understand the effect of the ring itself
on the particle motion to appreciate the more complex configurations
in which a ring is just part.

\section{The Dynamical System of Equations \label{DSE}}

Let us fix our coordinate system with axial symmetry: let $r\ge0$
be the coordinate away from the axis and $z\in\mathbb{R}$ the coordinate
along the axis. The dynamical system evolves with either a time coordinate
or a proper time. In any case a curve in the half-plane $r-z$ will
be parametrized by a {}``time'' parameter and its coordinate rate
of change denoted by $\dot{r}$, $\dot{z}$ and so on.

The dynamical system of equations will come either from the geodesic
equations in a spacetime or from Newton equations of motion of a test
particle in gravitational potential. We assume that both the spacetime
and the potential are static and axisymmetric. So our dynamical system
depends either on metric functions for geodesics or on the gravitational
potential for Newtonian motion.

Next we present the explicit equations to be solved.

\subsection{Geodesics in Weyl Solutions\label{GWS}}

The static spacetime of an axially symmetric body can be described
by the Weyl metric\begin{equation}
ds^{2}=e^{2\phi}\, dt^{2}-e^{-2\phi}\,\left[e^{2\gamma}\,\left(dr^{2}+dz^{2}\right)+r^{2}\, d\varphi^{2}\right]\label{weylmetric}\end{equation}
 where the functions $\phi$ and $\gamma$ depend only on $r$ and
$z$; the ranges of the coordinates $(r,z,\varphi)$ are the usual
for cylindrical coordinates and $t\in\mathbb{R}$. The vacuum Einstein's
equations ($R_{\mu\nu}=0$) reduce to the Laplace equation in cylindrical
coordinates, \begin{equation}
\phi_{,rr}+\frac{1}{r}\phi_{,r}+\phi_{,zz}=0\label{laplacian}\end{equation}
and the quadrature,\begin{equation}
d\gamma[\phi]=r\left[\left(\phi_{,r}^{\:2}-\phi_{,z}^{\:2}\right)\, dr+2\phi_{,r}\phi_{z}\, dz\right].\label{weylintegral}\end{equation}
 If $\phi$ satisfies the Laplace equation (\ref{laplacian}) then
$\gamma$ is twice differentiable. The function $\phi$ determines
the Weyl solution uniquely up to a constant.

The geodesic equations in Weyl spacetimes have two constants of motion
associated to the cyclic variables $t$ and $\varphi$, \begin{equation}
E=e^{2\phi}\dot{t},\quad L=r^{2}e^{-2\phi}\dot{\varphi}.\label{WeylConstMotion}\end{equation}
 Now the overdots mean derivation with respect to the proper time,
$\tau=s$. The other two second order evolution equations are

\begin{equation}
\ddot{r}=-(\dot{r}^{2}-\dot{z}^{2})(\gamma_{,r}-\phi_{,r})-2\dot{r}\dot{z}\left(\gamma_{,z}-\phi_{,z}\right)-P,\label{weylgeor}\end{equation}
\begin{equation}
\ddot{z}=(\dot{r}^{2}-\dot{z}^{2})(\gamma_{,z}-\phi_{,z})-2\dot{r}\dot{z}(\gamma_{,r}-\phi_{,r})-\phi_{,z}Q,\label{weylgeoz}\end{equation}
 where \[
P=e^{-2\gamma}\left[\phi_{,r}E^{2}+\left(\phi_{,r}-\frac{1}{r}\right)\frac{e^{4\phi}}{r^{2}}L^{2}\right],\]
 \[
Q=e^{-2\gamma}\left[E^{2}+\frac{e^{4\phi}}{r^{2}}L^{2}\right].\]

From the constant of motion $g_{\mu\nu}\,\dot{x}^{\mu}\dot{x}^{\nu}=1$
we find \begin{equation}
F(r,z)\equiv1+\frac{e^{2\phi}}{r^{2}}L^{2}-\frac{E^{2}}{e^{2\phi}}=-e^{2(\gamma-\phi)}\left(\dot{r}^{2}+\dot{z}^{2}\right).\label{weyleff}\end{equation}
 This function has some similarity with the Newtonian effective potential.
The motion is allowed only where $F(r,z)\le0$.

\subsection{Newtonian Motion in Axisymmetric Potential\label{NMASP}}

In Newtonian gravitation, the motion of a test particle in a field
of forces described by an axially symmetric potential $\phi$, solution
of Laplace equation, is characterized by two constants of motion:
the energy $H$ and the angular momentum $L$, \[
H=\frac{1}{2}\left(\dot{r}^{2}+r^{2}\dot{\varphi}^{2}+\dot{z}^{2}\right)+\,\phi(r,z),\]
\begin{equation}
L=r^{2}\dot{\varphi},\label{Ldef}\end{equation}
 where the over dots mean derivation with respect to the Newtonian
time; and two second order differential equations\begin{equation}
\ddot{r}=\frac{L^{2}}{r^{3}}-\;\phi_{,r},\qquad\ddot{z}=-\,\phi_{,z}.\label{newfallr}\end{equation}
 From these equations we have the constant of motion\[
H=\frac{1}{2}(\dot{r}^{2}+\dot{z}^{2})+V(r,z),\]
 where $V(r,z)$ is the effective potential, \begin{equation}
V(r,z)=\frac{L^{2}}{2r^{2}}+\,\phi(r,z).\label{neweff}\end{equation}
In the Newtonian case the motion is allowed where $V(r,z)\le H$.

\subsection{Axial and Equatorial Motions \label{sub:Axial-and-Equatorial}}

It is easy to see that if the problem has both axial and planar symmetry
at $z=0$ and no source except for the ring, then a test particle
with $z=0=\dot{z}$ is confined in the plane $z=0$ since the partial
derivatives of the functions along $z$ at $z=0$ is zero. The so
called equatorial motion.

There is also an axial motion. A test particle with $r=0=\dot{r}$
has vanishing angular momentum $L=0$ and by hypothesis of axial symmetry
and no source at the axis, the functions has vanishing partial derivative
along $r$ at $r=0$. Therefore $\ddot{r}=0$ and the particle stays
on the axis.

Of course the origin is an equilibrium point as long as the hypothesis
above are satisfied.

The general motion of a test particle can be very complicated for
non trivial functions. Care must be taken because the functions have
singularities at the ring.

\section{Ring Potential\label{RP}}

The ring we are concerned with uses the function $\phi$ which solves
the Laplacian (\ref{laplacian}) everywhere in the half plane $r-z$
except for the ring at $z=0,\, r=a$. It is the gravitational potential
itself for a Newtonian motion or the metric function which sets the
Weyl solution of the spacetime in which we have to solve the geodesic
motion. 

\begin{figure}

\caption{Contour plot of the potential $\phi$ for a constant mass density
ring with mass $M=1$ and radius $a=1$.\label{fig:wrphi}}

\includegraphics[%
  width=0.60\textwidth,
  keepaspectratio]{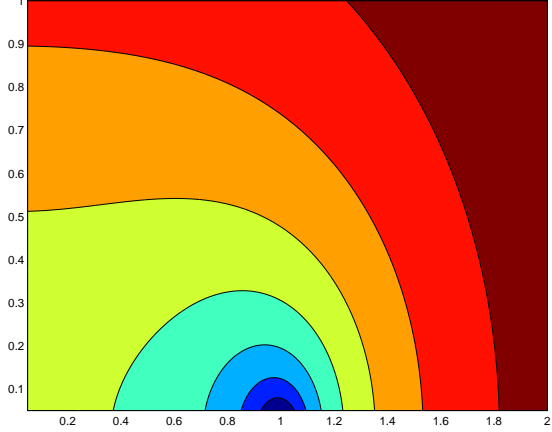}
\end{figure}

The Weyl-Bach solution has as Newtonian image the usual potential
for a ring of uniform density which is a solution of Laplace equation.
It can be written as \begin{equation}
\phi=-\frac{2M}{\pi R_{a}}\mathbf{K}\left(\frac{2\sqrt{ar}}{R_{a}}\right)=-\frac{M}{2\pi}\int_{0}^{2\pi}\frac{d\varphi}{\sqrt{r^{2}+z^{2}+a^{2}-2ar\,\cos\varphi}},\label{wrphi}\end{equation}
 where $R_{a}^{2}=(a+r)^{2}+z^{2}$ and $\mathbf{K}(x)$ is the complete
elliptic integral of the first kind for $x\in[0,1)$. The ring is
located on the plane $z=0$ and its center has coordinates $r=z=0$.
We shall take from now on $M=1=a$. The contour plot of this function
is depicted in Fig.~\ref{fig:wrphi}. 

The evaluation of the elliptic integral as well as its derivatives
was made with an algorithm adapted from the one presented in~\cite{NumericalR}.
Of special interest are the values of $\phi_{,r}$ at the axis $r=0$
and $\phi_{,z}$ at the plane $z=0$. Using the integral representation
above (\ref{wrphi}) we get\[
\lim_{r\rightarrow0}\phi_{,r}=\frac{M}{2\pi}\int_{0}^{2\pi}\frac{-a\cos\varphi\, d\varphi}{\left[z^{2}+a^{2}\right]^{3/2}}=0,\quad\forall z\]
and\[
\lim_{z\rightarrow0}\phi_{,z}=\frac{M}{2\pi}\int_{0}^{2\pi}\lim_{z\rightarrow0}\frac{z\, d\varphi}{\left[r^{2}+z^{2}+a^{2}-2ar\,\cos\varphi\right]^{3/2}}=\left\{ \begin{array}{lcl}
0 & \,\textrm{if}\, & r\ne a\\
-\infty & \,\textrm{if}\, & r\rightarrow a_{-}\\
+\infty & \,\textrm{if}\, & r\rightarrow a_{+}\end{array}\right.\]
Thus, the only Newtonian source for the potential is the ring itself.

\section{Newtonian Motion\label{sec:Newtonian-Motion}}

The equilibrium (circular) position of a test particle, if it exists,
obeys the equations $\dot{r}=\dot{z}=\ddot{r}=\ddot{z}=0$, that is,
from the effective potential (\ref{neweff})\[
\partial_{r}\phi=\frac{L^{2}}{r^{3}}\,,\quad\partial_{z}\phi=0.\]
 The planar symmetry implies the position may be in the $z=0$ plane.
And for the ring potential $\partial_{z}\phi>0$ for $z\ne0$. Thus
the motion is stable about the ring's plane for appropriate energies.

In the disk inside the ring we have $\partial_{r}\phi\le0$ so an
equilibrium point is possible only for $L=0$ at $r=0$ but it is
not stable since any amount of angular moment $L$ will push the particle
towards the inner part of the ring. 

If there is a velocity in the $z$ directions, $\dot{z}\ne0$, and
for small values of angular momentum, the test particle moves up and
down the disk inside the ring and for slightly higher angular momenta
it may cross in and out of the ring. 

For $L\ne0$, the centrifugal force pushes the particles away from
the axis. 

Outside the ring but in its plane, there is a lower bound of angular
momentum, let us say $L_{s}$, beyond which there are stable equilibrium
motions at a distance $r_{s}>a$. Using (\ref{neweff}) and $a=1=M$
we obtain $r_{s}$ and $L_{s}$ by setting $\partial V/\partial r=0=\partial^{2}V/\partial r^{2}$
at $z=0$. We find that $L_{s}=3.8396$ and $r_{s}=1.6095$.

\begin{figure}

\caption{Effective potential of the ring at $z=0,\, r=1$, for $r\in(0,3)$
and $z=0,\,\frac{1}{2},\,1$.\label{fig:wrnpef2}}

\subfigure[Case $L=L_s/8$.]{\includegraphics[%
  clip,
  width=0.30\textwidth,
  keepaspectratio]{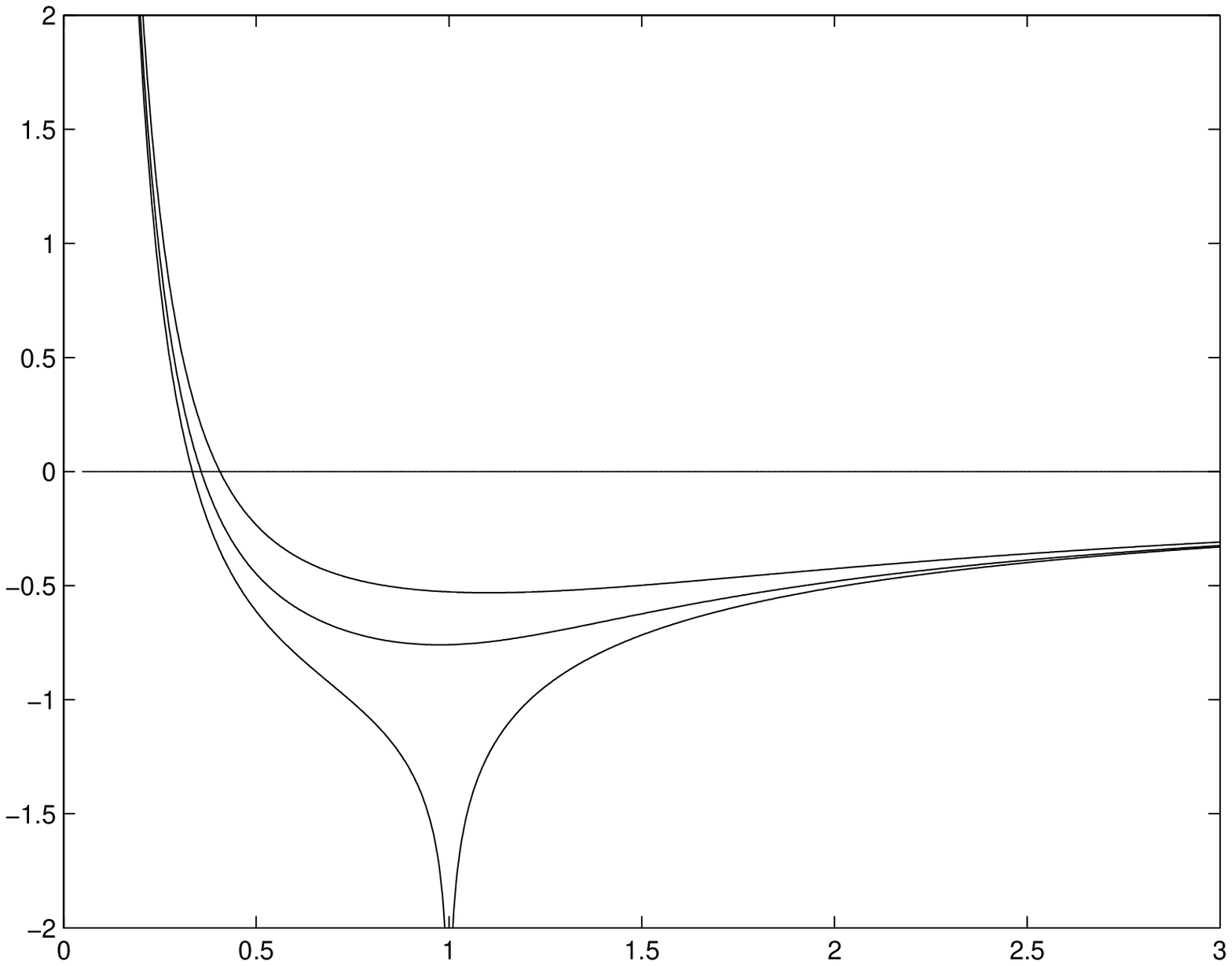}}\hfill{}\subfigure[Case $L=L_s/4$.]{\includegraphics[%
  clip,
  width=0.30\textwidth,
  keepaspectratio]{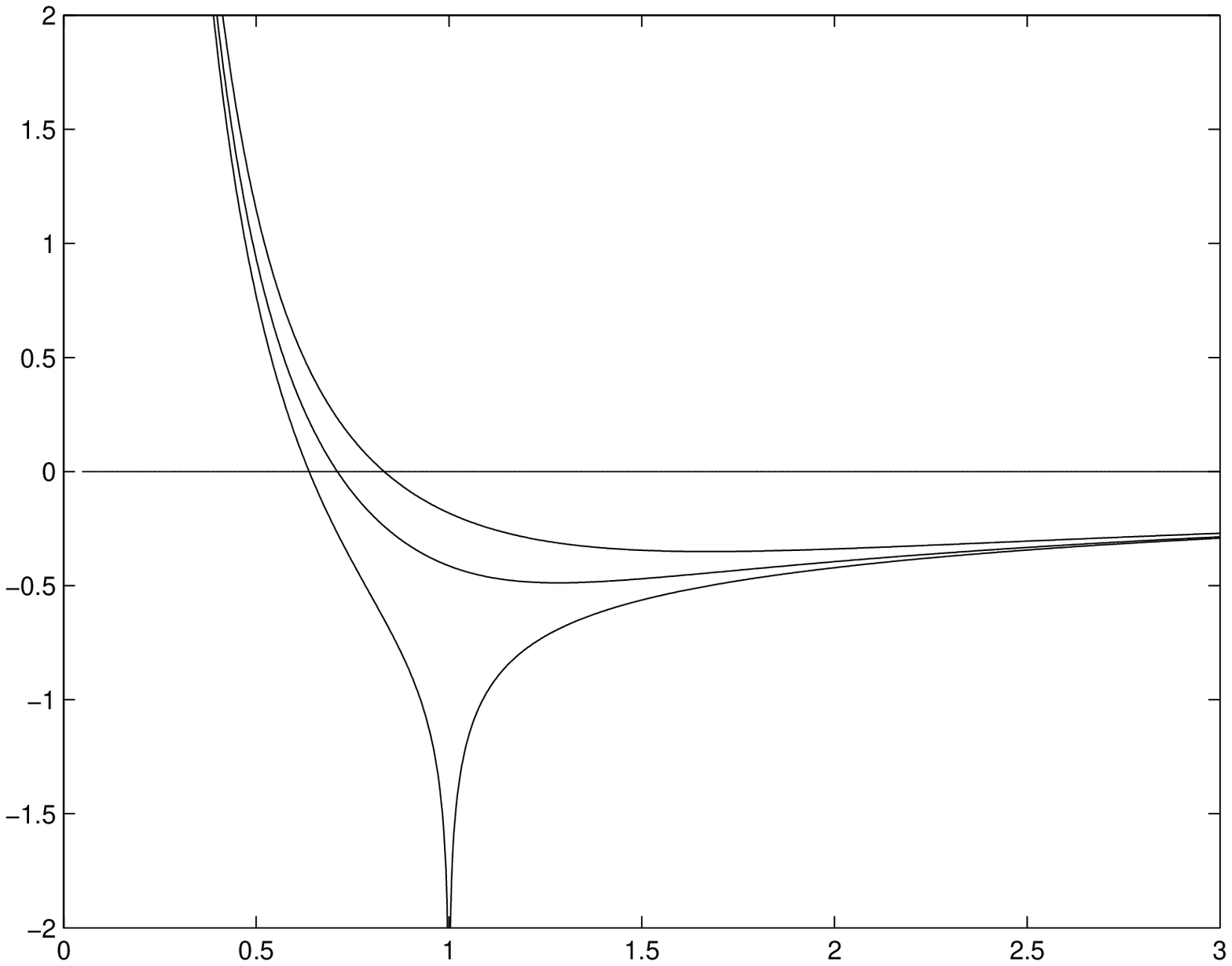}}\hfill{}\subfigure[Case $L=L_s/2$.]{\includegraphics[%
  clip,
  width=0.30\textwidth,
  keepaspectratio]{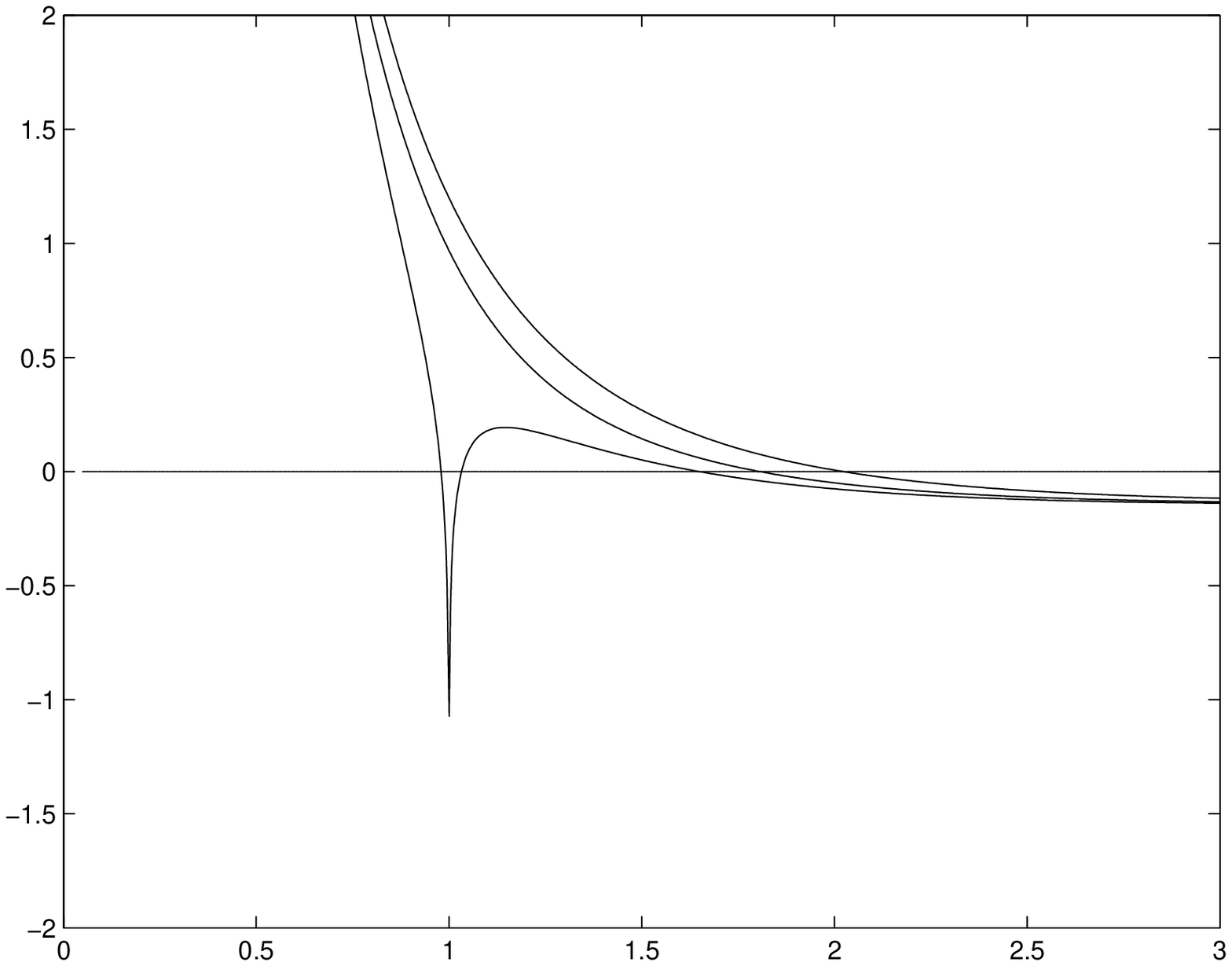}}
\end{figure}

We show in Figs.~ \ref{fig:wrnpef2}a,b,c the effective potential
for the ring $M=1=a$, for different values of angular momentum, $L=2^{-k}L_{s},$
for $k=3,\,2,\,1$ respectively.

\begin{figure}

\caption{Trajectory of a free falling particle in the gravitational field
of a ring of constant density. The particle initially at rest ($L=0$)
at $(r,z)=(1,1)$. \label{fig:wrnew1}}

\includegraphics[%
  clip]{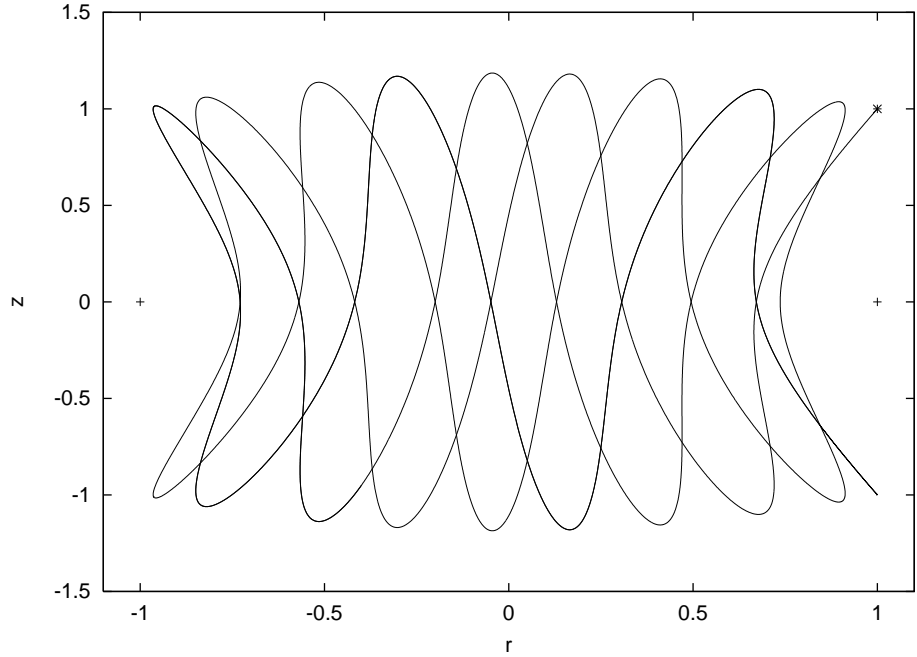}
\end{figure}

The motion of a test particle initially at rest at $r=z=1$ ( $L=0$)
is study in Fig.~\ref{fig:wrnew1}. %

\begin{figure}

\caption{Trajectory of a free falling particle with initial conditions $r=z=1$,
$\dot{r}=\dot{z}=0$.\label{fig:wrnew2}}

\subfigure[Case $L=1$.]{\includegraphics[%
  clip,
  width=0.50\textwidth,
  keepaspectratio]{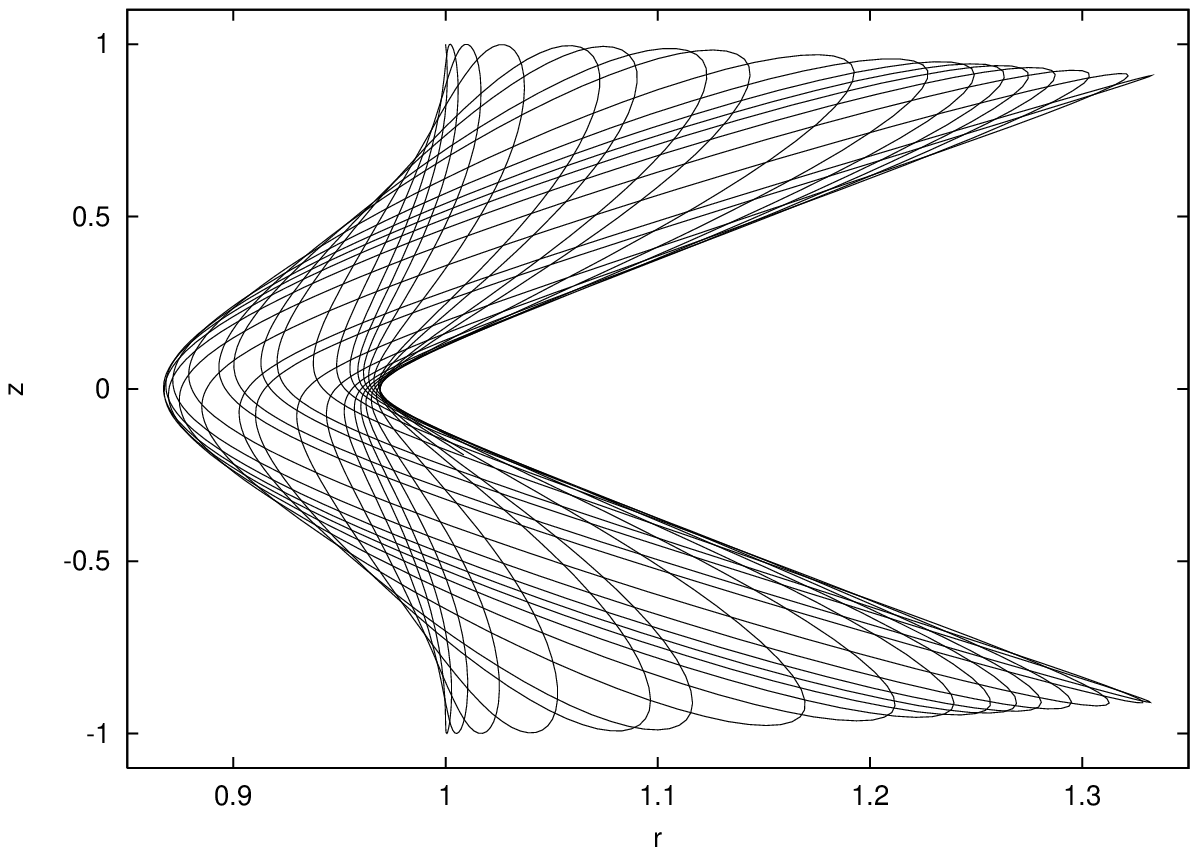}}\subfigure[Case $L=4.5$.]{\includegraphics[%
  clip,
  width=0.50\textwidth,
  keepaspectratio]{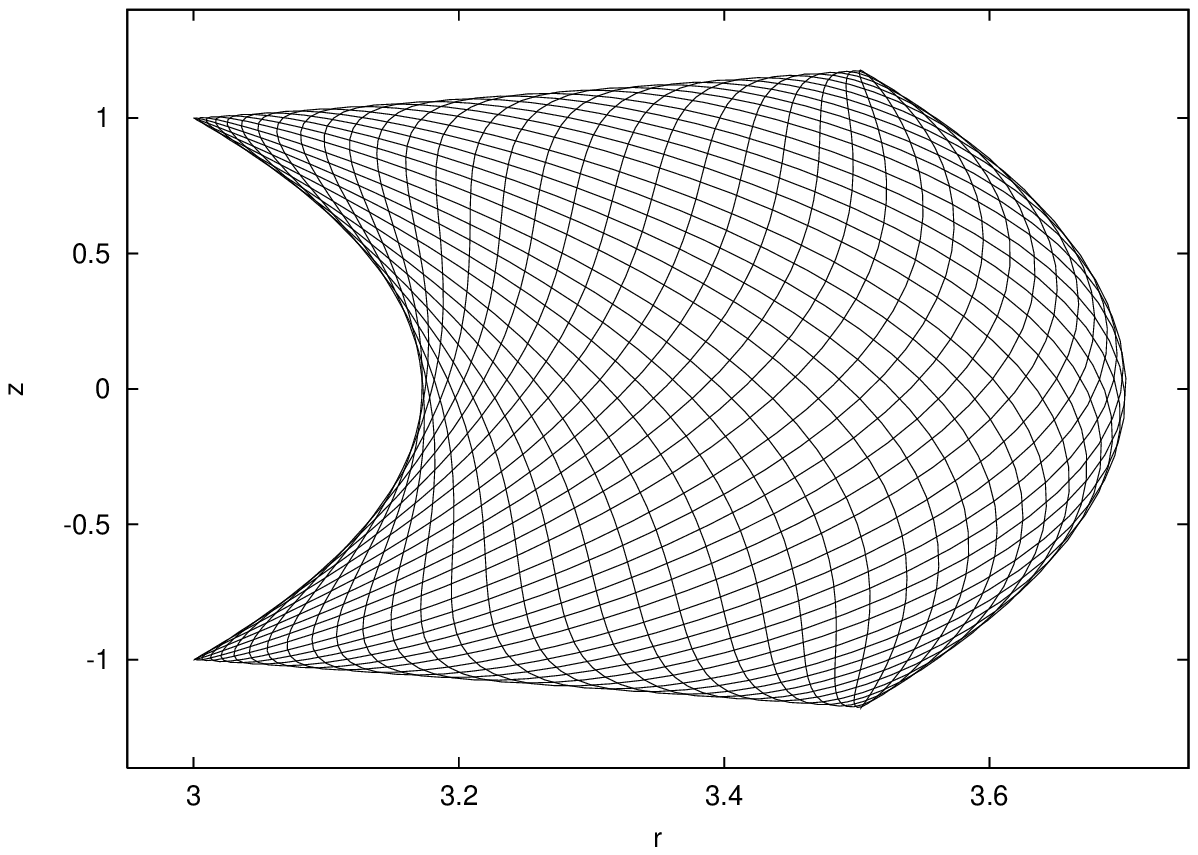}}
\end{figure}

In Figs.~\ref{fig:wrnew2}a,b we present trajectories with initial
conditions $r=z=1$, $\dot{r}=\dot{z}=0$ and different angular momenta,
$L=1$ and $L=4.5$, respectively. The orbits in these two cases are
bounded in a tri-dimensional region of the space. %

\section{Geodesics in Weyl-Bach Ring\label{sec:Geodesics-in-Weyl-Bach}}

Now for the general relativistic case we have two {}``potentials'',
$\psi$ and $\gamma$. The metric functions for the Weyl-Bach ring
are better expressed in toroidal coordinates $(\eta,\xi)\in[0,\infty)\times[0,2\pi)$
which are related to the axial coordinates $(r,z)$ by \[
r=a\frac{\sinh\eta}{\cosh\eta-\cos\xi},\quad z=a\frac{\sin\xi}{\cosh\eta-\cos\xi},\]
\[
\cot\xi=\frac{r^{2}+z^{2}-a^{2}}{2az},\quad\coth\eta=\frac{r^{2}+z^{2}+a^{2}}{2ar}\]

\begin{figure}

\caption{Contour plot of $\gamma$ for the Weyl-Bach solution associated to
a ring with $M=1=a$. \label{fig:wrnu}}

\includegraphics[%
  clip,
  width=0.60\textwidth,
  keepaspectratio]{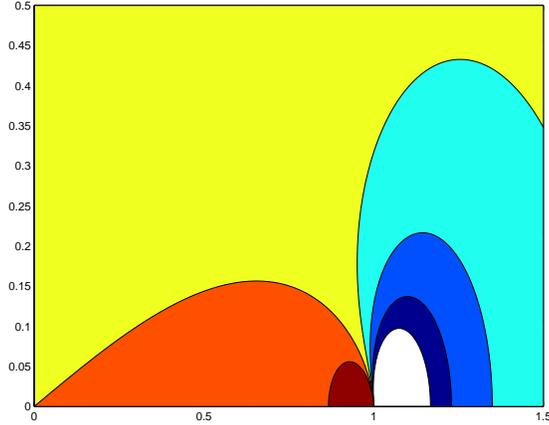}
\end{figure}

The Newtonian potential can be cast as \[
\phi=-\sigma e^{-\eta/2}\mathbf{K}(\kappa)\sqrt{\cosh\eta-\cos\xi},\]
where $\mathbf{K}(\kappa)$ is the complete elliptic integral of first
kind, $\kappa^{2}=1-e^{-2\eta}$ and $\sigma\equiv\frac{\sqrt{2}M}{\pi a}$.
Then the $\gamma$ function for the Weyl-Bach solution is \cite{Weyl(1922),Hoenselaers(1995),Semerak},
\begin{equation}
\gamma=-\frac{\sigma^{2}}{2}\mathbf{K}\left[\mathbf{K}\left\{ 1+\kappa'^{2}-\kappa^{2}\left(2+\kappa'^{2}\right)\frac{\cos\xi}{\sinh\eta}\right\} -2\mathbf{E}\left\{ 1-\kappa^{2}\frac{\cos\xi}{\sinh\eta}\right\} \right]\label{eq:gamWB}\end{equation}
where $\mathbf{E}$ is the complete elliptic integral of second kind
and $\kappa'=e^{-\eta}$. For the actual computations of the function
$\gamma$ this formula is not particularly interesting. We found more
convenient its evaluation by direct integration of equation (\ref{weylintegral})
using Gauss quadrature methods for chosen integration paths. The contour
of the function~(\ref{eq:gamWB}) is shown in Fig.~\ref{fig:wrnu}.

\begin{figure}

\caption{Geodesics in Weyl-Bach ring solution with initial conditions $(r,z)=(2,0)$,
$\dot{r}=0$ and some values of $E$ with $a=1=\sigma$. \label{fig:wrgeo}}

\includegraphics[%
  clip,
  width=0.60\textwidth,
  keepaspectratio]{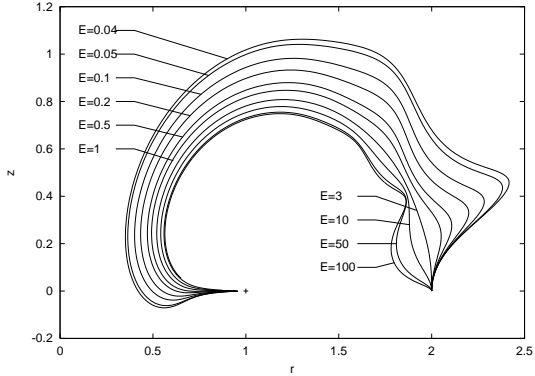}
\end{figure}

In Fig.~\ref{fig:wrgeo} we display time-like geodesics of test particles
moving in the gravitational field of Weyl-Bach ring. The initial conditions
are $(r,z)=(2,0)$ with initial velocity only in the $z$ direction
determined by the values of $E$ shown in the graphic and $L=0$.
For low values of $E$, we have a clear repulsion at the beginning
of the trajectories. %
 The repulsion happens a little later for higher values of $E$. The
geodesic with $E=3$ seems to suffer only attraction. Therefore we
have a separatrix closer to this value of $E$. 

These geodesics have interesting asymptotic behavior. They approach
the ring from inside its interior at very high values of the proper
time, that is, the Weyl-Bach ring has a directional singularity and
the particle hits the ring only at infinite proper time. This behavior
seems to be generic. 

The behavior of these geodesics is characterized by the following
two features: a) All these geodesics are asymptotically `radial'.
That is, there is a privileged directions around the ring, pointing
to a directional singularity of the Riemann tensor similar to the
singularity of the Chazy-Curson metric~\cite{Letelier(1998)}. The
curvature scalar invariants $w_{1}=\frac{1}{8}C_{abcd}C^{abcd}$ and
$w_{2}=\frac{1}{16}C_{ab}^{cd}C_{cd}^{ef}C_{ef}^{ab}$, where $C^{abcd}$
is the Weyl tensor have different limits when one approaches the singularity
from different directions~\cite{Letelier(1998)}. Both invariants
blow without bound when the limit is taken from the interior of the
ring and approach zero from up or down directions. b) We have the
`freezing' of the motion in Weyl coordinates. The numerical computation
of these geodesics for very high values of proper time indicates that
they will reach the singularity at an infinite proper time.

\begin{figure}

\caption{Geodesics in the Weyl-Bach ring solution starting at $(r,z)=(2,0)$
with $E=1$ and several values of $L$. \label{fig:wrL}}

\includegraphics[%
  clip,
  width=0.80\textwidth,
  keepaspectratio]{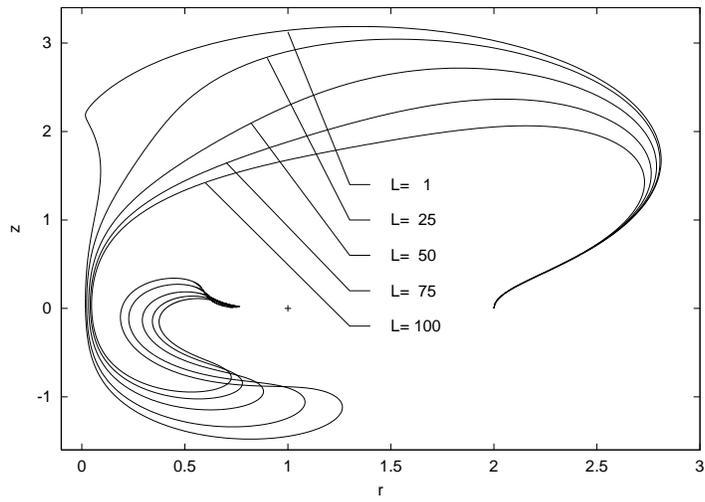}
\end{figure}

\begin{figure}

\caption{Geodesics of Weyl-Bach ring solution starting at rest from $(r,z)=(1,1),\,(2,1)$,
$(2.00,0.01),\,(2.0,-0.5)$. All the geodesics reach the ring singularity
from the inner disk. \label{fig:wrfree}}

\includegraphics[%
  clip,
  width=0.80\textwidth,
  keepaspectratio]{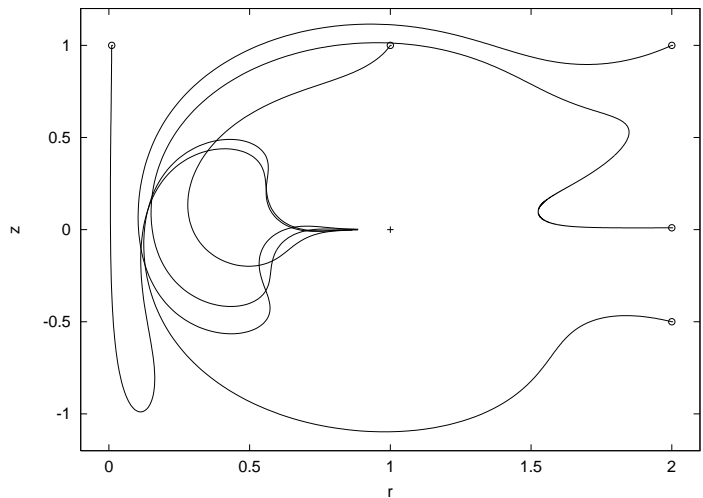}
\end{figure}

In Fig.~\ref{fig:wrL} we plot geodesics initially at $(r,z)=(2,0)$,
with $E=1$, initial velocity with no component in the $r$ direction,
and $L=1,25,50,75,$ and $100$. All the geodesics finish in the ring
by its inner disk at infinite proper time. %
Finally, in Fig.~\ref{fig:wrfree} we present five trajectories of
test particles initially at rest at $(r,z)=(1,1),(2,1),(2,0.01),$
and $(2,-0.5).$ The geodesic that starts at $(2,0.01)$ suffers a
very strong repulsion. Again, all the geodesics fall in the ring exactly
as the previous case. %

It is clear the structure of the inner part of the ring is non trivial.
Any geodesic approaching the inner part of the ring has either infinite
proper time or infinite proper length. One can compute a proper distance
to the ring of a curve in the $(r,z)$ plane, in which the ring is
at $(a,0)$:\begin{equation}
\left|\;\int\limits _{(r,z)}^{(a,0)}e^{\gamma-\phi}\sqrt{dr^{2}+dz^{2}}\,\right|.\end{equation}
 Let us show that it diverges according to the different directions
of the approach using the toroidal coordinates $(\eta,\xi)$ toroidal
coordinates. As $\eta\rightarrow\infty$ and $\xi\in(\frac{\pi}{2},\frac{3\pi}{2})$
the approach is from inside. Using the asymptotic of the Elliptic
functions one can show that\[
\lim_{(r,z)\rightarrow(a,0)}(\gamma-\phi)=-\lim_{\eta\rightarrow\infty}\left[\left(\frac{M}{a\pi}\right)^{2}\cosh\eta\,\cos\xi\right].\]
 Thus the ring is at finite distance when approached from outside
(in which $\cos\xi\ge0$) because the integrand goes to zero, whereas
it is at infinite distance when approached from inside (in which $\cos\xi<0$)
because the integrand diverges. Therefore a fixed \emph{physical}
distance to the ring means greater (smaller) \emph{coordinate} distance
in the directions where $\lim_{\Sigma\rightarrow0}(\gamma-\phi)$
is small (large). Hence, the particles should appear {}``repelled''
({}``attracted'') in the directions from where the ring is physically
nearby (far away). This agrees with what can really be seen in the
figures.

\section{Discussion}

We investigated the gravitation induced by a ring both in the Newtonian
and in General Relativity dynamics of test particles. Although related,
the spacetime associated to the Newtonian potential of a ring has
quite distinct features. We have learned that line sources \cite{Israel66}
in General Relativity exhibits directional singularities and the results
above for the Weyl-Bach ring give explicit example of them.

Furthermore, the Weyl coordinates have the tendency to compact a whole
region into a singularity, This is the case for the Schwarzschild
solution in which the event horizon and the physical singularity (with
the topology of $S^{2}\times\mathbb{R}$ and $\mathbb{R}$ respectively)
are displayed in Weyl conformal coordinates as a single world finite
line segment ($I\times\mathbb{R}$ where $I\subset\mathbb{R}$ is
a line segment).

And we find very interesting that the inner side of the ring is very
attractive but is not accessible for particle geodesics because it
is too far away. The test particles approach the singularity in a
privileged way: They arrive along radial directions of the ring inner
disk. We presented geodesics which take an infinite amount of proper
time to hit the ring. This happens also in the extreme case of the
Reissner-Nordstrom solution. The physical distance to the event horizon
is infinite.

The geodesics display gravitational field with apparent repulsive
regions. This can be either a coordinate effect as pointed out above
or may indicate the presence of very high tensions in the ring. Probably
both. 

The apparition of tensions in the Weyl solutions is a known fact \cite{Letelier(1998)}.
The imposition of a static geometry and the Einstein's equation creates
some devices like strings, struts, membranes, etc. to support an otherwise
dynamical configuration. In the present case we have some kind of
strong hoop tension along the ring. As we know the spacetime is sensitive
to both density and pressure or tension while in the Newtonian gravitation
the density of the source suffices for the gravitational potential.

So far, there are few self-gravitating ring solutions of Einstein's
equations \cite{Kodama(2003)}. The reader should be cautious about
rings solutions in the literature \cite{Appell(1887),Gleiser(1989),Letelier(1987),Letelier(1998)}.
Some have misprints, others have misinterpretation (see Appendix A).
Nevertheless they are very interesting.

In this paper we show some interesting behavior of test particles
about a ring alone. We think the understanding of the gravitation
of the ring itself is useful for configurations in which the ring
is an important part.

We thank FAPESP and CNPq for financial support and Eduardo Gueron
for the critical reading.

\section*{A. Rings Problems in the Literature}

In 1887 a very simple solution of Poisson equation which is singular
at a ring was published \cite{Appell(1887)}. And indeed it was interpreted
as a ring-like configuration of matter. This interpretation is present,
at least, in two important books \cite{Bateman,WhittakerWatson}.
Based on that, Letelier and Oliveira \cite{Letelier(1987)} developed
a series of new potentials with singularities at the ring and were
misinterpreted as generated by rings. Moreover they were able to get
a Weyl class of new solutions for Einstein equations in vacuum with
axial symmetry. Nevertheless, as Gleiser and Pullin \cite{Gleiser(1989)}
correctly showed, Appell solution is not just a ring. Actually there
is a surface mass density in the plane of the ring. Let us compute
it because this is the source of another mistake caused by a misprint.

The Appell potential is\[
\phi=-\frac{M}{\sqrt{r^{2}+(z-A)^{2}}}\]
where $r$ and $z$ are standard cylindrical coordinates and the constants
$M$ and $A$ may take complex values. Physical potential is the real
part of $\phi$. We promptly see the singularity at $r=a,\, z=0$
if $A=ia$, that is, it is singular at the ring. Nevertheless the
$z$ derivative is not null at $z=0$. Actually there is a jump across
this plane:\[
\lim_{z\rightarrow0}\left[\phi_{,z}\right]=-2MA\left(r^{2}+A^{2}\right)^{-\frac{3}{2}}.\]
If one takes the real part of the potential with $A=ia$, where $a$
and $M$ are positive real constants one gets the surface mass density:\[
\sigma=\left\{ \begin{array}{cc}
-4Ma\left(a^{2}-r^{2}\right)^{-\frac{3}{2}} & \quad\textrm{for}\quad r<a.\\
0 & \quad\textrm{for}\quad r>a.\end{array}\right.\]
In \cite{Gleiser(1989)} the power is misprinted as $-\frac{1}{2}$.
It happens that there is a known disk with such a surface mass density,
in the class of the so called Morgan \& Morgan family \cite{MorganMorgan}
of disks in General Relativity.

This mistake lead us into another misinterpreted result. If both disks
had the same surface mass density one could subtract one from the
other leaving just the ring \cite{Letelier(1998)}. But Appell disk
(sic) and Morgan \& Morgan disk do not have the same surface mass
density!

Of course, looking back, one could not have more than one ring as
solution of Laplace equation with axial symmetry. There exist several
theorems proving the existence and uniqueness of solution of the Laplace
equation. For an infinitesimal ring with axial symmetry, there is
no possibility of other solution but the constant linear mass density.
And the Weyl solution linked to the potential is also unique. This
is the ring of the main part of this paper with the corresponding
Weyl Bach solution spacetime.

\end{document}